\documentclass[11pt]{article} 

\usepackage{helvet}

\newcommand*{\myEXPfont}{\fontfamily{cmr}\selectfont}

\usepackage{graphicx} 
\usepackage{mathtools}
\usepackage{amsmath}
\usepackage{multirow}
\usepackage[export]{adjustbox}
\usepackage[left=1in, right=1in, top=1in, bottom=1in]{geometry} 
\usepackage{booktabs}
\usepackage{rotating}
\usepackage{rotfloat}
\usepackage{float}
\usepackage{subfig}
\usepackage{url}
\usepackage{pdfpages}
\usepackage{rotating}
\usepackage[doublespacing]{setspace} 
\usepackage{csvsimple}
\usepackage{array}
\usepackage{wrapfig}
\usepackage[utf8]{inputenc}

\usepackage{cite}
\usepackage{tabularx}
\usepackage{float}

\usepackage{amsthm}                
\usepackage{amssymb}

\usepackage{multirow}

\usepackage{caption}

\usepackage{amsfonts}
\usepackage{amsmath} 
\usepackage{tikz}
\usetikzlibrary{positioning,shapes.geometric}

\usepackage{xcolor}

\usepackage{setspace}

\DeclareMathOperator{\E}{\mbox{{\myEXPfont E}}}

\usepackage{mathrsfs}

\makeatletter
\newcommand*{\indep}{%
  \mathbin{%
    \mathpalette{\@indep}{}%
  }%
}
\newcommand*{\nindep}{%
  \mathbin{
    \mathpalette{\@indep}{\not}
  }%
}
\newcommand*{\@indep}[2]{%
  \sbox0{$#1\perp\m@th$}
  \sbox2{$#1=$}
  \sbox4{$#1\vcenter{}$}
  \rlap{\copy0}
  \dimen@=\dimexpr\ht2-\ht4-.2pt\relax
  \kern\dimen@
  {#2}%
  \kern\dimen@
  \copy0 
} 
\makeatother

\usepackage{xr-hyper}

\definecolor{forestgreen}{RGB}{34,139,34}
\usepackage{hyperref}
\hypersetup{
    colorlinks=true,
    linkcolor=magenta,
    filecolor=magenta, 
    citecolor=forestgreen,      
    urlcolor=blue
}

\usepackage{sectsty}
\sectionfont{\large}

\usepackage{layout}

\usepackage{geometry}

\newcolumntype{C}[1]{>{\centering\arraybackslash}p{#1}}

\usepackage{authblk}

\usepackage{afterpage}
\usepackage{lscape}
\usepackage{graphicx}

\usepackage{datetime}
\def\paperversionmajor{3}
\def\paperversionminor{0}

\makeatletter
\newcommand*{\addFileDependency}[1]{
  \typeout{(#1)}
  \@addtofilelist{#1}
  \IfFileExists{#1}{}{\typeout{No file #1.}}
}
\makeatother

\begin{document}

\title{Learning about treatment effects in a new target population under transportability assumptions for relative effect measures}

\author[1-3]{Issa J. Dahabreh}
\author[1,2]{Sarah E. Robertson}
\author[4]{Jon A. Steingrimsson}

\affil[1]{CAUSALab, Harvard T.H. Chan School of Public Health, Boston, MA}
\affil[2]{Department of Epidemiology, Harvard T.H. Chan School of Public Health, Boston, MA}
\affil[3]{Department of Biostatistics, Harvard T.H. Chan School of Public Health, Boston, MA}
\affil[4]{Department of Biostatistics, Brown University School of Public Health, Providence, RI }

\maketitle{}
\thispagestyle{empty} 

\newpage 
\thispagestyle{empty} 

\vspace{0.1in}
\noindent \textbf{Address for correspondence:} Dr. Issa J. Dahabreh, Department of Epidemiology, Harvard T.H. Chan School of Public Health, Boston, MA 02115; email: \href{mailto:idahabreh@hsph.harvard.edu}{idahabreh@hsph.harvard.edu}; phone: +1 (617) 495‑1000. 

\vspace{0.1in}
\noindent \textbf{Type of manuscript:} Original research article.

\vspace{0.1in}
\noindent \textbf{Running head:} Learning with transportability assumptions for relative effect measures

\vspace{0.1in}
\noindent \textbf{Conflicts of interest/Competing interests:} The authors have no relevant financial or non-financial interests to disclose.

\vspace{0.1in}
\noindent \textbf{Funding:} This work was supported in part by National Library of Medicine (NLM) Award R01LM013616, and Patient-Centered Outcomes Research Institute (PCORI) awards ME-1502-27794 and ME-2019C3-17875.

\vspace{0.1in}
\noindent \textbf{Author contribution:} All authors were involved in drafting the manuscript and have read and approved the final version submitted. No statistical analyses are included in the paper.

\vspace{0.1in}
\noindent \textbf{Data and computing code availability:} Not applicable. 

\vspace{0.1in}
\noindent \textbf{Consent to Participate/Publish (Ethics):} This research did not involve human subjects.

\vspace{0.1in}
\noindent
\textbf{Word count:} abstract = 255; main text $\approx$ 3725.

\vspace{0.1in}
\noindent
\textbf{Abbreviations:} No abbreviations used.

\newpage
\thispagestyle{empty} 


\vspace*{1in}
{\LARGE \centering Learning about treatment effects in a new target population under transportability assumptions for relative effect measures \par }

\vspace{1in}
\noindent
\textbf{Running head:} Learning with transportability assumptions for relative effect measures

\vspace{0.3in}
\noindent
\textbf{Type of manuscript:} Original Research Article.

\vspace{0.3in}
\noindent
\textbf{Word count:} abstract = 255; main text $\approx$ 3725.

\vspace{0.3in}
\noindent
\textbf{Abbreviations that appear in the text:} No abbreviations used.


\clearpage
\thispagestyle{empty} 
\vspace*{0.75in}
\begin{abstract}
\noindent
\linespread{1.3}\selectfont
Epidemiologists and applied statisticians often believe that relative effect measures conditional on covariates, such as risk ratios and  mean ratios, are ``transportable'' across populations. Here, we examine the identification of causal effects in a target population using an assumption that conditional relative effect measures (e.g., conditional risk ratios or mean ratios) are transportable from a trial to the target population. We show that transportability for relative effect measures is largely incompatible with transportability for difference effect measures, unless the treatment has no effect on average or one is willing to make even stronger transportability assumptions, which imply the transportability of both relative and difference effect measures. We then describe how marginal causal estimands in a target population can be identified under the assumption of transportability of relative effect measures, when we are interested in the effectiveness of a new experimental treatment in a target population where the only treatment in use is the control treatment evaluated in the trial. We extend these results to consider cases where the control treatment evaluated in the trial is only one of the treatments in use in the target population, under an additional partial exchangeability assumption in the target population (i.e., a partial assumption of no unmeasured confounding in the target population). We also develop identification results that allow for the covariates needed for transportability of relative effect measures to be only a small subset of the covariates needed to control confounding in the target population. Last, we propose estimators that can be easily implemented in standard statistical software. 
\end{abstract}

\clearpage
 \pagenumbering{arabic}
\setcounter{page}{1}
\section*{INTRODUCTION}

Many epidemiologists and applied statisticians believe that relative effect measures, such as risk ratios and mean ratios, possibly conditional on baseline covariates, are ``transportable'' across populations or at least more likely to be transportable compared with absolute effect measures \cite{glasziou1995evidence, schwartz2006ratio, spiegelman2017evaluating}. For example, it is often believed that the (conditional) relative risk estimated in a trial is likely to apply to a new population of individuals in which the randomized treatments may be considered for clinical use. Yet, prior work on transportability methods has focused on assumptions of transportability of difference effect measures \cite{dahabreh2020transportingStatMed}, or stronger assumptions requiring exchangeability in expectation or in distribution between the trial and the target population (e.g., \cite{pearl2015, dahabreh2020transportingStatMed}). The only exception we are aware of is the work by Huitfeld et al. \cite{huitfeldt2019effect}, which formally described an assumption of transportability of relative effect measures, but did not develop complete identification results using that assumption.

Here, we show that the condition of transportability of relative effect measures is incompatible with the previously described condition of transportability of difference effect measures, unless the treatment has no effect on average or one is willing to accept even stronger transportability assumptions (that imply transportability of both relative and difference effect measures). Under the assumption of transportability of relative effect measures, we develop identification results for the causal mean ratio and the average treatment effect in the target population. We begin by considering identification when the only treatment in use in the target population is the control treatment that is evaluated in the trial. We then extend these results to allow for treatment variation in the target population, such that the control treatment is only one of multiple treatments in use in the target population, under a partial exchangeability condition (i.e., a no unmeasured confounding assumption for the control treatment in the target population). We also develop identification results that allow for the covariates needed for transportability of relative effect measures to be only a small subset of the covariates needed to control confounding in the target population. Last, we describe easy to implement estimators, including some that do not require the availability of outcome information from the target population but instead use externally developed or ``transported'' models (from a different source population).

\section*{STUDY DESIGN, NOTATION, AND CAUSAL ESTIMANDS} \label{section_design_data_estimand}

\paragraph{Study design:} Consider a non-nested study design where the investigators have data from a randomized trial that compared the treatments of interest and a separately obtained sample from a target population in which experimentation is not feasible \cite{dahabreh2021studydesigns}. This design can be used to learn about treatment effects in a target population that meets the eligibility criteria of the trial (in generalizability analyses) as well as broader target populations (in transportability analyses) \cite{dahabreh2019commentaryonweiss}. Because we expect that the most common application of the proposed methods will be in transportability analyses, we use that term throughout. Our results, however, apply to both generalizability and transportability analyses. 

\paragraph{Data:} Let $X$ denote the baseline covariates; $S$ the source population ($S = 1$ for trial participants; $S = 0$ for the target population); $A$ the treatment; and $Y$ the binary, count, or continuous outcome measured at the end of the study. For simplicity, we assume that the treatment is binary. We shall refer to $A = 1$ as the experimental treatment and $A = 0$ as the control treatment. We model the trial data as independent and identically distributed draws of $(X_i, S_i = 1, A_i, Y_i)$, $i = 1, \ldots, n_1$, where $n_1$ is the number of trial participants. We model the data from the target population as independent and identically distributed draws of $(X_i, S_i = 0, A_i, Y_i)$, $i = 1, \ldots, n_0$, where $n_0$ is the number of individuals in a target population sample. We define $n$ as the total sample size in the composite dataset that includes data from both the trial and the target population sample, that is, $n = n_0 + n_1$. The distribution of the observable variables $(X, S, A, Y)$ in the population underlying the trial may be different from the distribution in the target population. For example, the baseline covariates have a different distribution in the two populations reflecting selective participation in the trial. Furthermore, the conditional distribution of treatment given the baseline covariates is also likely to differ between the two populations (e.g., treatment is often marginally randomized in $S=1$ but depends on measured or unmeasured variables in $S = 0$). Throughout, we use $f(\cdot)$ to generically denote densities. In what follows all densities and expectations are with respect to the sampling scheme underlying the non-nested trial design \cite{dahabreh2021studydesigns}.

\paragraph{Causal estimands:} To define the causal estimands of interest, let $Y^a$ denote the potential (counterfactual) outcome under intervention to set treatment $A$ to $a$, possibly contrary to fact \cite{splawaneyman1990, rubin1974, robins2000d}. Here, $a = 1$ denotes intervention to set $A$ to the experimental treatment; $a=0$ denotes intervention to set $A$ to the control treatment. 

In the non-nested trial design, without additional knowledge regarding the marginal or conditional probability of participation in the trial, tenable estimands of interest are those in the target population (of non-randomized individuals) \cite{dahabreh2019commentaryonweiss}. Here, we will focus on learning about the marginal causal mean ratio in the target population, $$\dfrac{\E[Y^1 | S = 0 ]}{\E[Y^0 | S = 0 ]}, \mbox{ with } \E[Y^0 | S = 0 ] \neq 0.$$ When the outcome is binary, the causal mean ratio equals the causal risk ratio  in the target population, $\dfrac{\Pr[Y^1 = 1 | S = 0 ]}{\Pr[Y^0 = 1 | S = 0 ]}$. We use the more general term ``mean ratio'' because our results also apply to continuous and count outcomes. For completeness, we will also provide identification results for the average treatment effect in the target population: $$ \E[Y^1 - Y^0 | S = 0 ] = \E[Y^1 | S = 0 ] - \E[Y^0 | S = 0 ].$$

\subsection*{UNIFORM USE OF THE CONTROL TREATMENT IN THE TARGET POPULATION} 

We begin by considering the situation where the experimental treatment evaluated in the trial is unavailable in the target population and the control treatment evaluated in the trial is the only treatment in use in the target population. In other words, we assume that $A = 0$ for all individuals in the target population. This may be the case if the trial compares a new experimental treatment before approval for clinical use (a setup also adopted in \cite{huitfeldt2018choice}), against a control treatment that is uniformly applied in the target population (i.e., a ``standard-of-care'' treatment).

\subsection*{Identifiability conditions}

We shall use the following identifiability conditions: 

\noindent
\emph{(A1) Consistency of potential outcomes:} if $A_i = a$, then $Y_i = Y^{a}_i$, for each $a \in \{0, 1\}$ and for every individual $i$.

\noindent
\emph{(A2) Mean exchangeability over $A$ in the trial:} for each $a \in \mathcal \{0,1\}$ and every $x$ with positive density in the trial $f(x, S = 1 ) \neq 0$, $\E [ Y^{a} | X = x , S = 1, A =a] = \E [ Y^{a} | X = x, S = 1]$.

\noindent
\emph{(A3) Positivity of treatment assignment in the trial:} $\Pr[A=a | X = x, S=1] > 0$ for each $a \in \{0,1\}$ and each $x$ with positive density in the trial $f(x , S = 1) \neq 0$.

\noindent
\emph{(A4) Conditional transportability (exchangeability) of relative effect measures over $S$ \cite{huitfeldt2019effect}:} 
\begin{equation*}
    \begin{split}
        &\dfrac{\E [ Y^1 | X = x , S = 1]}{\E [ Y^0 | X = x , S = 1]} = \dfrac{\E [ Y^1  | X = x , S = 0]}{\E [ Y^0 | X = x , S = 0]}, \\
   &\quad\quad\quad \E [ Y^0 | X = x , S = 1] \neq 0 \mbox{ and } \E [ Y^0 | X = x , S = 0] \neq 0 ,
    \end{split}
\end{equation*}
for every $x$ with positive density in the target population $f(x, S = 0 ) \neq 0$. 

\noindent
\emph{(A5) Uniform use of the control treatment in the target population:} $S = 0 \implies A = 0$, that is, if $S_i = 0$, then $A_i = 0$ for every individual $i$.

\noindent
\emph{(A6) Positivity of trial participation:} $\Pr[S=1 | X = x] >0,$ for every $x$ such that $f(x, S = 0 ) \neq 0.$

\paragraph{Reasoning about the identifiability conditions:} Conditions (A1) through (A3), and (A6), are often invoked in generalizability or transportability analyses and we refer readers to recent work discussing them in detail \cite{dahabreh2019identification, dahabreh2020transportingStatMed}. 

Condition (A4), the key identifiability condition in this paper, connects conditional relative effect measures in the trial and the target population. This condition is often assumed informally when interpreting the results of trials and was formally stated in \cite{huitfeldt2019effect}. The positivity components of the formal statement of the condition are mild technical requirements to preclude division by zero. Leaving the positivity components aside, condition (A4) is weaker than assumptions of exchangeability in expectation or in distribution between the trial and the target population (see, e.g., \cite{cole2010, pearl2015, dahabreh2018generalizing, rudolph2017, dahabreh2020transportingStatMed}). 
Here, by ``weaker'' we mean that the assumptions of exchangeability  in expectation or in distribution between the trial and the target population imply, but are not necessarily implied, by condition (A4).

\subsection*{Relationship between transportability of relative and difference effect measures} 

In previous work (e.g., \cite{dahabreh2020transportingStatMed}), which aimed to identify the average treatment effect in the target population (i.e., to identify $\E[Y^1 - Y^0 | S = 0]$), when seeking to avoid assumptions of exchangeability in expectation or in distribution between the trial and target populations, investigators have invoked the following condition, instead of condition (A4):

\emph{(A4$^*$) Conditional transportability (exchangeability) of difference effect measures over $S$ \cite{dahabreh2020transportingStatMed}:} 
\begin{equation*}
    \begin{split}
        \E [ Y^1 | X = x , S = 1] - \E [ Y^0 | X = x , S = 1] = \E [ Y^1  | X = x , S = 0] - \E [ Y^0 | X = x , S = 0] \\
    \end{split}
\end{equation*}
for every $x$ with positive density in the target population $f(x, S = 0 ) \neq 0$.

As is the case for transportability of relative effect measure, transportability of difference effect measures is weaker than exchangeability, in expectation or in distribution, between the trial and the target population. Thus, it is natural to wonder what is the relationship between the two conditions of transportability of effect measures. In the Appendix, we use a simple algebraic argument to show that if one assumes that both conditions (A4) and (A4$^*$) simultaneously hold, then at least one of the following implications has to be true, for all $x$ such that $f(x, S = 0) \neq 0$:  
\begin{enumerate}
    \item[I1.] the treatment has no effect on average, conditional on covariates, in both the trial and the target population, that is, it has to be that $ \E [ Y^1 | X = x , S = s] = \E [ Y^0 | X = x , S = s] $, for $s = 0, 1$; or
    \item[I2.] the potential outcomes are exchangeable in expectation over $S$, for both treatments, that is, it has to be that $ \E [ Y^a | X = x , S = 1] = \E [ Y^a  | X = x , S = 0] $, for $a = 0, 1$.
\end{enumerate}

It is also worth noting that if conditions (A1) through (A3), (A5), and (A6) also hold, then from the two implications above we can obtain the following restrictions on the law of the observed data, for all $x$ such that $f(x, S = 0) \neq 0$:
\begin{enumerate}
    \item[R1.] in the trial, the conditional expectation of the outcome within covariate level  $X = x$ among individuals who receive the experimental treatment has to equal the conditional expectation of the outcome within covariate level  $X = x$ among individuals who receive the control treatment, that is, $ \E [ Y | X = x , S = 1, A = 1] = \E [ Y | X = x , S = 1, A = 0] $; or
    \item[R2.] the conditional expectation of the outcome within covariate level  $X = x$ among individuals in the trial assigned to the control treatment has to equal the conditional expectation of the outcome within covariate level  $X = x$ among individuals in the target population, that is, $ \E [ Y | X = x , S = 1, A = 0] = \E [ Y  | X = x , S = 0] $.
\end{enumerate}
Interestingly, restrictions R1 and R2 are at least in principle testable using the observed data, though formal (statistical hypothesis) testing of these conditions is challenging when $X$ is high-dimensional. 

In practical applications, implication I1 will often be implausible. That is, we typically do not expect that the treatment has no effect (on average) within all levels of the covariates, both in the trial and the target population. Furthermore, methods that rely on transportability of (relative or difference) effect measures are often motivated by the belief that implication I2 is not plausible. In other words, reliance on assumptions (A4) or (A4$^*$) is typically most useful when exchangeability in expectation between the trial and the target population is not plausible. If we find implications I1 and I2 implausible, or if restrictions R1 and R2 are incompatible with the data, we should conclude that at least one of conditions (A4) and (A4$^*$) fails to hold (or that other conditions are violated).  

This analysis suggests that we should largely view transportability of relative effect measures as incompatible with transportability of difference effect measures. Thus, in practical applications, we have to rely on substantive background knowledge to decide which, if any, of the two conditions is more plausible. Arguably, epidemiological and statistical lore takes condition (A4) as more plausible than assumption (A4$^*$). Nevertheless, there is no logical reason to believe that is the case in general. We recommend case-by-case consideration of both assumptions in applied work. In the remainder of this paper, we will assume transportability of relative effect measures to obtain identification results for the causal estimands of interest.

\subsection*{Identification under uniform use of the control treatment in the target population} 

In the Appendix, we show that under conditions (A1) through (A6), listed above, the causal mean ratio $\dfrac{\E[Y^1 | S = 0]}{\E[Y^0 | S = 0]}$ is identifiable by 
\begin{equation} \label{eq:id_CRR_phi}
    \phi \equiv \dfrac{\E \left[ \dfrac{\E[Y | X, S = 1, A = 1]}{\E[Y | X, S = 1, A = 0]} \E[Y | X, S = 0] \Big | S = 0 \right]}{\E[Y | S = 0]}.
\end{equation}
Furthermore, the average treatment effect in the target population is identifiable by
\begin{equation*} 
    \beta \equiv \E \left[ \dfrac{\E[Y | X, S = 1, A = 1]}{\E[Y | X, S = 1, A = 0]} \E[Y | X, S = 0] \Big | S = 0 \right] - \E[Y | S = 0].
\end{equation*}

Detailed derivations for these results are given in the Appendix. Here, we just highlight that the results depend on condition (A4), which states that conditional relative effect measures are transportable (used to re-express the numerator of the causal mean ratio in terms of observable variables), \emph{as well as} condition (A5), which states that the control treatment is the only treatment available in the target population (used to re-express both the numerator and the denominator of causal mean ratio in terms of observable variables). 

\section*{TREATMENT VARIATION IN THE TARGET POPULATION} 

Thus far, we have worked under a setup where all members of the target population receive the control treatment $A = 0$. We now consider the situation where there exists treatment variation in the target population, such that the control treatment is not the only treatment in the target population. In that case, the identification results in the previous section are unlikely to apply; we provide details in the Appendix, but the key issue is that in the presence of treatment variation in the target population condition (A5) is violated and the identification results for the previous section can be invalidated by the presence of confounding in the target population.

Progress may be possible, however, if substantive knowledge suggests the following exchangeability condition in the target population. Specifically, consider the following two conditions:

\vspace{0.1in}
\noindent
\emph{(B1) Partial exchangeability in expectation over $A$ in the target population:} for every $x$ with positive density in the target population $f(x, S = 0 ) \neq 0$, $\E [ Y^0 | X = x , S = 0, A = 0] = \E [ Y^0 | X = x, S = 0]$.

\vspace{0.1in}
\noindent
\emph{(B2) Positivity of the conditional probability of receiving the control treatment $A=0$ in the target population:} for every $x$ with positive density in the target population $f(x, S = 0 ) \neq 0$, the probability of receiving the control treatment is positive $\Pr[A = 0 | X = x, S = 0] >0$.

\vspace{0.1in}
We refer to condition (B1) as ``partial exchangeability'' because it pertains to only counterfactuals under the control treatment (i.e., the intervention to set treatment $A$ to $a=0$). In applications, this condition will be an untestable (partial) assumption of no confounding in the target population; thus, it will require case-by-case examination on the basis of background knowledge. Similarly, condition (B2) only pertains to the target population probability of receiving the control treatment evaluated in the trial. This positivity condition is in principle testable but formal evaluation may be challenging if $X$ is high-dimensional \cite{petersen2012diagnosing}.

As we show in the Appendix, under conditions (A1) through (A4), (A6), (B1), and (B2), we can identify the causal mean ratio in the target population by 
\begin{equation} \label{eq:id_CRR_chi}
    \chi \equiv \dfrac{\E \left[ \dfrac{\E[Y | X, S = 1, A = 1]}{\E[Y | X, S = 1, A = 0]} \E[Y | X, S = 0, A = 0] \Big | S = 0 \right]}{ \E \big[  \E[Y | X, S = 0, A = 0]  \big | S = 0 \big]}.
\end{equation}
Furthermore, we can identify the average treatment effect in the target population by 
\begin{equation*} 
    \gamma \equiv \E \left[ \dfrac{\E[Y | X, S = 1, A = 1]}{\E[Y | X, S = 1, A = 0]} \E[Y | X, S = 0, A = 0] \Big | S = 0 \right] -  \E \big[  \E[Y | X, S = 0, A = 0]  \big | S = 0 \big].
\end{equation*}

\subsection*{What if confounding control in the target population requires additional covariates?}

To simplify the results above, we assumed that the same vector of covariates $X$ is adequate for condition (A4) and condition (B1). It will often be more reasonable, however, to assume that the set of covariates to render the trial and target population exchangeable is smaller than the set of covariates needed to control confounding in the target population. In fact, the preference for using relative effect measures is sometimes justified from the empirical observation that strong statistical interaction on the multiplicative scale is uncommon \cite{spiegelman2017evaluating}. Suppose then that we collect from the sample of the target population an additional set of baseline covariates $W$. In this situation, the data from the trial are modeled as independent and identically distributed draws of $(X_i, S_i = 1, A_i, Y_i)$, $i = 1, \ldots, n_1$, where $n_1$ is the number of trial participants, and the data from the target population are modeled as independent and identically distributed draws of $(X_i, W_i, S_i = 0, A_i, Y_i)$, $i = 1, \ldots, n_0$, where $n_0$ is the number of individuals in a target population sample. As before, we define $n = n_0 + n_1$.

Suppose that conditioning on $(X, W)$, but not $X$ alone, is sufficient to control confounding in the target population. This means that we replace the two conditions given in the preceding section, with the following ones:

\vspace{0.1in}
\noindent
\emph{(C1) Partial mean exchangeability over $A$ in the target population:} for every $x, w$ with positive density in the target population $f(x, w, S = 0 ) \neq 0$, $\E [ Y^0 | X = x, W = w, S = 0, A = 0] = \E [ Y^0 | X = x, W = w, S = 0]$.

\vspace{0.1in}
\noindent
\emph{(C2) Positivity of the conditional probability of receiving the control treatment $A=0$ in the target population:} for every $x, w$ with positive density in the target population $f(x,w, S = 0 ) \neq 0$, the probability of receiving the control treatment is positive $\Pr[A = 0 | X = x, W = w, S = 0] >0$.

In the Appendix, we show that under conditions (A1) through (A4), (A6), (C1), and (C2) the mean ratio in the target population is identified by
\begin{equation} \label{eq:id_CRR_psi}
    \psi \equiv \dfrac{\E \left[ \dfrac{\E[Y | X, S = 1, A = 1]}{\E[Y | X, S = 1, A = 0]}  \E \big[ \E[Y | X, W, S = 0, A = 0] \big | X, S = 0 \big] \Big |  S = 0 \right] }{ \E \big[  \E[Y | X, W, S = 0, A = 0]  \big | S = 0 \big]}.
\end{equation}
Furthermore, we show that the average treatment effect in the target population is identifiable by
\begin{equation*}
    \begin{split}
        \delta &\equiv \E \left[ \dfrac{\E[Y | X, S = 1, A = 1]}{\E[Y | X, S = 1, A = 0]}  \E \big[ \E[Y | X, W, S = 0, A = 0] \big | X, S = 0 \big] \Big |  S = 0 \right] \\ 
        &\quad\quad\quad\quad\quad\quad\quad\quad\quad\quad - \E \big[  \E[Y | X, W, S = 0, A = 0]  \big | S = 0 \big].
    \end{split}
\end{equation*}

\section*{ESTIMATION \& INFERENCE} 

\paragraph{Estimation:} The most straightforward approach to estimate the functionals in the preceding section is to replace all quantities in the identification results with their sample analogs. To be concrete, starting with the causal mean ratio results, for $\phi$ we might use
\begin{equation}\label{eq:estimation_phi}
    \widehat\phi = \dfrac{ \sum_{i=1}^n (1 - S_i) \widehat r(X_i) \widehat g (X_i) }{\sum_{i=1}^n (1 - S_i) Y_i },
\end{equation}
where $\widehat r(X) $ is an estimator for the conditional relative effect measure in the trial, $\dfrac{\E[Y | X, S = 1, A = 1]}{\E[Y | X, S = 1, A = 0]}$; and $\widehat g (X) $ is an estimator for the conditional expectation of the outcome in the target population, $\E[Y | X, S = 0]$. For $\chi$ we might use
\begin{equation}\label{eq:estimation_chi}
    \widehat\chi = \dfrac{ \sum_{i=1}^n (1 - S_i) \widehat r(X_i) \widehat h (X_i) }{\sum_{i=1}^n (1 - S_i) \widehat h (X_i) },
\end{equation}
where $\widehat r(X)$ is as defined above and $\widehat h (X)$ is an estimator for $\E[Y | X, S = 0, A = 0]$.
Last, for $\psi$ we might use 
\begin{equation}\label{eq:estimation_psi}
    \widehat\psi = \dfrac{ \sum_{i=1}^n (1 - S_i) \widehat r(X_i) \widehat b (X_i) }{\sum_{i=1}^n (1 - S_i) \widehat m (X_i, W_i) },
\end{equation}
where $\widehat r(X)$ is as defined above, $\widehat m (X, W)$ is an estimator for $ \E[Y | X, W, S = 0, A = 0]$, and $\widehat b (X)$ is an estimator of $\E \big[ \E[Y | X, W, S = 0, A = 0] \big | X, S = 0 \big]$. For completeness, in the Appendix we give analogous results for estimating the identifying functionals for the average treatment effect. 

Implementing the proposed estimators is relatively straightforward: for example, for a binary outcome $Y$, we can use relative risk regression to estimate $\widehat r(X)$ and any kind of regression appropriate for binary outcomes to estimate $\widehat g (X)$, $\widehat h(X)$, or $\widehat m(X, W)$. To estimate $\widehat b (X)$ we have to resort to a slightly more complicated procedure, inspired by similar methods for the iterated expectation formulation of the g-formula in observational analyses: (1) use a standard regression approach to obtain $\widehat m(X, W)$ by regressing the outcome $Y$ on $(X, W)$ among untreated individuals in the target population sample; (2) obtain predictions by applying $\widehat m(X, W)$ to the entire sample from the target population (regardless of treatment); (3) use a regression (e.g., ordinary least squares regression) of the predictions from the previous step on $X$ alone, to obtain  $\widehat b (X)$. The terms $\widehat b(X_i)$ in formula \eqref{eq:estimation_psi} can then obtained by evaluating the fitted model from step (3) at the observed $X_i$ value of each observation in the target population sample.

Because $X$ and (especially) $W$ may be fairly high dimensional, estimation of $\widehat r(X) $, $\widehat g (X)$, $\widehat h (X)$, $\widehat m(X, W)$, and  $\widehat b (X)$ will typically require the specification of parametric or semiparametric models \cite{robins1997toward}. If the models are correctly specified and converge to the true underlying conditional expectations at parametric rate, then the estimators will be consistent. Estimation using data-adaptive methods that converge at slower than parametric rate may be the subject of future work \cite{chernozhukov2018double}.

\paragraph{Inference:} Inference for estimators of the identifying functionals (e.g., for the mean ratio estimators $\widehat \phi$, $\widehat \chi$, and $\widehat \psi$) can be carried out using standard (stacked) M-estimation methods \cite{stefanski2002} to obtain ``sandwich'' estimators of sampling variances, accounting for uncertainty in the estimation of the models the estimators depend on. Alternatively, the non-parametric bootstrap \cite{efron1994introduction} or other simulation-based methods \cite{greenland2004interval} can be used.

\paragraph{Using externally developed or ``transported'' models:} In the Appendix, we discuss versions of the estimators given above that do not require outcome information from the target population, but instead use externally developed or transported models (i.e., models developed using external datasets but tailored for the target population \cite{steingrimsson2021transportingmodels}). As we explain in the Appendix, these approaches are less than ideal because they require additional transportability assumptions for outcome regression models; thus, when investigators have outcome data from the target population the methods described in the should be used instead. That said, using externally developed or transported models may be useful when the outcome is not measured in the target population and well-calibrated (to the target population) outcome models are available from external sources or can be developed using previously described methods \cite{shimodaira2000improving, sugiyama2012machine, steingrimsson2021transportingmodels}.

\section*{DISCUSSION} 

We examined the identification of marginal causal mean ratios and average treatment effects under assumptions of conditional transportability of relative effect measures from a trial to a target population. Our results, which involve standardization of relative effect measures to the target population \cite{miettinen1972standardization, greenland1982interpretation}, are appropriate when (1) the control treatment evaluated in the trial is the only treatment in use in the target population, or (2) there exists variation in treatment, but a partial exchangeability assumption is plausible for the target population. Informally, both of these situations can be thought of as indicating the absence of confounding in the target population, either due to lack of variation in treatment or because the measured baseline covariates are adequate for confounding control. In the presence of variation in treatment in the target population, we provided identifiability results that allow the covariates needed for the assumption of transportability of relative effect measures to be only a (possibly small) subset of the covariates needed for confounding control in the target population. These results should prove useful when substantive knowledge suggests that there are few relative effect measure modifiers and confounding control in the target population requires adjustment for many additional covariates (i.e., confounders). 

The condition of transportability of relative effect measures \cite{huitfeldt2019effect}, which underpins all our results, is similar in spirit to transportability of difference effect measures \cite{dahabreh2020transportingStatMed}, in the sense that both conditions are weaker than exchangeability in expectation or in distribution between the trial and the target population. We argued that the conditions of transportability of relative and difference effect measures are largely incompatible. The practical implication is that we have to rely on scientific judgment to select which, if any, of these two conditions can be reasonably assumed to hold in practical applications. Interestingly, a similar conundrum arises in observational studies using rate change methods, where investigators have to decide whether the rate change in the control group is a good proxy for the counterfactual rate change in the treated group had it been untreated, on the difference or ratio scale \cite{van2021causal}. Furthermore, we showed that, the simultaneous adoption of the conditions of transportability in relative and difference measures, when combined with other identifiability conditions, imposes testable restrictions on the law of the observable data. Though we discussed these restrictions under a setup where condition (A5) holds, analogous restrictions can be obtained after replacing condition (A5) with conditions (B1) and (B2), or (C1) and (C2). 

When outcome data are hard to obtain from the target population, versions of our methods that circumvent the need for such data by using externally developed or transported outcome regression models may be useful. Combining a ``baseline risk model'' with a ``relative risk'' estimate from a randomized trial is a common approach in decision and economic analyses. Thus, our results can be viewed as a formalization of the conditions needed to endow such analyses with a causal interpretation. Of note, in decision and economic analyses the relative measure of effect obtained from the trial and used for modeling is often marginal (i.e., population-averaged); reference \cite{hong2019comparison} gives a recent epidemiological example using this approach. Thus, it would appear that investigators are sometimes willing to make the strong assumption that the marginal relative effect measure in the trial is transportable to the target population. In our approach, the transportability assumption is only made within levels of baseline covariates (as was the case in \cite{huitfeldt2019effect}), which may be more plausible because it allows for treatment effect heterogeneity over the covariates. 

Though weaker than other commonly invoked assumptions, transportability of relative effect measures is still a fairly strong assumption. Future work may consider the development of sensitivity analysis methods to allow examination of the potential impact of unmeasured effect modifiers of the relative effect measure. Future work may also address incomplete adherence to treatment and consider extensions to failure time outcomes.

In summary, we have presented methods for learning about treatment effects in a new target population under assumptions of conditional transportability of relative effect measures between a trial and the target population. These methods may be a useful addition to the toolbox of investigators extending inferences from randomized trials to target populations of substantive interest, particularly when background knowledge suggests that relative effect measures are modified by a relatively small number of measured baseline covariates.

\section*{APPENDIX MATTERIAL}

All Appendices are availabe at \href{https://www.dropbox.com/s/rw0balcnwxh1chx/APPENDIX_transportability_of_relative_effect_measures.pdf?dl=0}{appendix link}.

\section*{DECLARATIONS}

\noindent


\noindent
\textbf{Availability of data and material:} Not applicable.

\noindent
\textbf{Code availability:} Not applicable.

\clearpage
\bibliographystyle{ieeetr}
\bibliography{bibliography_cate}

\begin{thebibliography}{10}

\bibitem{glasziou1995evidence}
P.~P. Glasziou and L.~M. Irwig, ``An evidence based approach to individualising
  treatment,'' {\em BMJ}, vol.~311, no.~7016, pp.~1356--1359, 1995.

\bibitem{schwartz2006ratio}
L.~M. Schwartz, S.~Woloshin, E.~L. Dvorin, and H.~G. Welch, ``Ratio measures in
  leading medical journals: structured review of accessibility of underlying
  absolute risks,'' {\em BMJ}, vol.~333, no.~7581, p.~1248, 2006.

\bibitem{spiegelman2017evaluating}
D.~Spiegelman and T.~J. VanderWeele, ``Evaluating public health interventions:
  6. modeling ratios or differences? let the data tell us,'' {\em American
  Journal of Public Health}, vol.~107, no.~7, pp.~1087--1091, 2017.

\bibitem{dahabreh2020transportingStatMed}
I.~J. Dahabreh, S.~E. Robertson, J.~A. Steingrimsson, E.~A. Stuart, and M.~A.
  Hern{\'a}n, ``Extending inferences from a randomized trial to a new target
  population,'' {\em Statistics in Medicine}, vol.~39, no.~14, pp.~1999--2014,
  2020.

\bibitem{pearl2015}
J.~Pearl, ``Generalizing experimental findings,'' {\em Journal of Causal
  Inference}, vol.~3, no.~2, pp.~259--266, 2015.

\bibitem{huitfeldt2019effect}
A.~Huitfeldt, S.~A. Swanson, M.~J. Stensrud, and E.~Suzuki, ``Effect
  heterogeneity and variable selection for standardizing causal effects to a
  target population,'' {\em European Journal of Epidemiology}, vol.~34, no.~12,
  pp.~1119--1129, 2019.

\bibitem{dahabreh2021studydesigns}
I.~J. Dahabreh, S.~J.-P. Haneuse, J.~M. Robins, S.~E. Robertson, A.~L.
  Buchanan, E.~A. Stuart, and M.~A. Hern\'an, ``Study designs for extending
  causal inferences from a randomized trial to a target population,'' {\em
  American Journal of Epidemiology}, vol.~190, no.~8, pp.~1632--1642, 2021.

\bibitem{dahabreh2019commentaryonweiss}
I.~J. Dahabreh and M.~A. Hern{\'a}n, ``Extending inferences from a randomized
  trial to a target population,'' {\em European Journal of Epidemiology},
  vol.~34, no.~8, pp.~719--722, 2019.

\bibitem{splawaneyman1990}
J.~Splawa-Neyman, ``On the application of probability theory to agricultural
  experiments. essay on principles. section 9. [{T}ranslated from
  {S}plawa-{N}eyman, {J} (1923) in {R}oczniki {N}auk {R}olniczych {T}om {X},
  1--51],'' {\em Statistical Science}, vol.~5, no.~4, pp.~465--472, 1990.

\bibitem{rubin1974}
D.~B. Rubin, ``Estimating causal effects of treatments in randomized and
  nonrandomized studies.,'' {\em Journal of {E}ducational {P}sychology},
  vol.~66, no.~5, pp.~688--701, 1974.

\bibitem{robins2000d}
J.~M. Robins and S.~Greenland, ``Causal inference without counterfactuals:
  comment,'' {\em Journal of the American Statistical Association}, vol.~95,
  no.~450, pp.~431--435, 2000.

\bibitem{huitfeldt2018choice}
A.~Huitfeldt, A.~Goldstein, and S.~A. Swanson, ``The choice of effect measure
  for binary outcomes: Introducing counterfactual outcome state transition
  parameters,'' {\em Epidemiologic Methods}, vol.~7, no.~1, 2018.

\bibitem{dahabreh2019identification}
I.~J. Dahabreh, J.~M. Robins, S.~J.-P. Haneuse, and M.~A. Hern\'an,
  ``Generalizing causal inferences from randomized trials: counterfactual and
  graphical identification,'' {\em arXiv preprint arXiv:1906.10792}, 2019
  (accessed: 11/03/2020).

\bibitem{cole2010}
S.~R. Cole and E.~A. Stuart, ``Generalizing evidence from randomized clinical
  trials to target populations: the {A}{C}{T}{G} 320 trial,'' {\em American
  {J}ournal of {E}pidemiology}, vol.~172, no.~1, pp.~107--115, 2010.

\bibitem{dahabreh2018generalizing}
I.~J. Dahabreh, S.~E. Robertson, E.~J. Tchetgen~Tchetgen, E.~A. Stuart, and
  M.~A. Hern{\'a}n, ``Generalizing causal inferences from individuals in
  randomized trials to all trial-eligible individuals,'' {\em Biometrics},
  vol.~75, no.~2, pp.~685--694, 2018.

\bibitem{rudolph2017}
K.~E. Rudolph and M.~J. van~der Laan, ``Robust estimation of encouragement
  design intervention effects transported across sites,'' {\em Journal of the
  Royal Statistical Society. Series B (Statistical Methodology)}, vol.~79,
  no.~5, pp.~1509--1525, 2017.

\bibitem{petersen2012diagnosing}
M.~L. Petersen, K.~E. Porter, S.~Gruber, Y.~Wang, and M.~J. van~der Laan,
  ``Diagnosing and responding to violations in the positivity assumption,''
  {\em Statistical Methods in Medical Research}, vol.~21, no.~1, pp.~31--54,
  2012.

\bibitem{robins1997toward}
J.~M. Robins and Y.~Ritov, ``Toward a curse of dimensionality appropriate
  ({C}{O}{D}{A}) asymptotic theory for semi-parametric models,'' {\em
  Statistics in Medicine}, vol.~16, no.~3, pp.~285--319, 1997.

\bibitem{chernozhukov2018double}
V.~Chernozhukov, D.~Chetverikov, M.~Demirer, E.~Duflo, C.~Hansen, W.~Newey, and
  J.~Robins, ``Double/debiased machine learning for treatment and structural
  parameters,'' {\em The Econometrics Journal}, vol.~21, no.~1, pp.~C1--C68,
  2018.

\bibitem{stefanski2002}
L.~A. Stefanski and D.~D. Boos, ``The calculus of {M}-estimation,'' {\em The
  American Statistician}, vol.~56, no.~1, pp.~29--38, 2002.

\bibitem{efron1994introduction}
B.~Efron and R.~J. Tibshirani, {\em An introduction to the bootstrap}.
\newblock No.~57 in Monographs on Statistics and Applied Probability, Boca
  Raton, Florida, USA: Chapman \& Hall/CRC, 1993.

\bibitem{greenland2004interval}
S.~Greenland, ``Interval estimation by simulation as an alternative to and
  extension of confidence intervals,'' {\em International Journal of
  Epidemiology}, vol.~33, no.~6, pp.~1389--1397, 2004.

\bibitem{steingrimsson2021transportingmodels}
J.~A. Steingrimsson, C.~Gatsonis, and I.~J. Dahabreh, ``Transporting a
  prediction model for use in a new target population,'' {\em arXiv preprint
  arXiv:2101.11182}, 2021.

\bibitem{shimodaira2000improving}
H.~Shimodaira, ``Improving predictive inference under covariate shift by
  weighting the log-likelihood function,'' {\em Journal of Statistical Planning
  and Inference}, vol.~90, no.~2, pp.~227--244, 2000.

\bibitem{sugiyama2012machine}
M.~Sugiyama and M.~Kawanabe, {\em Machine learning in non-stationary
  environments: introduction to covariate shift adaptation}.
\newblock MIT press Cambridge Massachusetts, 2012.

\bibitem{miettinen1972standardization}
O.~S. Miettinen, ``Standardization of risk ratios,'' {\em American Journal of
  Epidemiology}, vol.~96, no.~6, pp.~383--388, 1972.

\bibitem{greenland1982interpretation}
S.~Greenland, ``Interpretation and estimation of summary ratios under
  heterogeneity,'' {\em Statistics in Medicine}, vol.~1, no.~3, pp.~217--227,
  1982.

\bibitem{van2021causal}
R.~van Aalst, E.~Thommes, M.~Postma, A.~Chit, and I.~J. Dahabreh, ``On the
  causal interpretation of rate-change methods: the prior event rate ratio and
  rate difference,'' {\em American Journal of Epidemiology}, vol.~190, no.~1,
  pp.~142--149, 2021.

\bibitem{hong2019comparison}
J.-L. Hong, M.~Webster-Clark, M.~Jonsson~Funk, T.~St{\"u}rmer, S.~E. Dempster,
  S.~R. Cole, I.~Herr, and R.~LoCasale, ``Comparison of methods to generalize
  randomized clinical trial results without individual-level data for the
  target population,'' {\em American Journal of Epidemiology}, vol.~188, no.~2,
  pp.~426--437, 2019.

\end{thebibliography}


\ddmmyyyydate 
\newtimeformat{24h60m60s}{\twodigit{\THEHOUR}.\twodigit{\THEMINUTE}.32}
\settimeformat{24h60m60s}
\begin{center}
\vspace{\fill}\ \newline
\textcolor{black}{{\tiny $ $generalizability\_relative\_measures, $ $ }
{\tiny $ $Date: \today~~ \currenttime $ $ }
{\tiny $ $Version: \paperversionmajor.\paperversionminor $ $ }}
\end{center}

\end{document}